\documentclass[12pt]{article}
\usepackage{epsf}
\usepackage{graphics}
\usepackage{color}
\usepackage{pazha}
\tightenlines

\def\*{$^{*}$}
\def\a{$^{\mbox{\footnotesize a}}$}
\def\bb{$^{\mbox{\footnotesize b}}$}
\def\cc{$^{\mbox{\footnotesize c}}$}
\def\dd{$^{\mbox{\footnotesize d}}$}
\def\e{$^{\mbox{\footnotesize e}}$}
\def\f{$^{\mbox{\footnotesize f}}$}
\def\g{$^{\mbox{\footnotesize g}}$}
\def\h{$^{\mbox{\footnotesize h}}$}

\sloppypar

\begin{document}
\begin{flushleft}
{\it to be published in Astronomy Letters, 2017, v. 43, n. 9,
  pp. 583-594}\\
{\it (in Russian: Pis'ma v Astronomicheskii Zhurnal, 2017,
  v. 43, No. 9, pp. 643-654)}\\ [30mm]
\end{flushleft}

\baselineskip 21pt

\title{\bf MULTIPLE X-RAY BURSTS AND THE MODEL OF A SPREADING
  LAYER OF ACCRETING MATTER OVER THE NEUTRON STAR SURFACE}

\author{\bf S. A.~Grebenev\affilmark{*}, I. V.~Chelovekov}   

\affil{{\it Space Research Institute, Russian Academy of
    Sciences, \\
    Profsoyuznaya ul. 84/32, Moscow, 117997 Russia}}

\vspace{2mm}
\received{September 26, 2016}

\vspace{2mm}
\noindent
We report the detection during the \mbox{JEM-X/INTEGRAL}
observations of several X-ray bursters of series of close type I
X-ray bursts consisting of two or three events with a recurrence
time much shorter than the characteristic (at the observed mean
accretion rate) time of matter accumulation needed for a
thermonuclear explosion to be initiated on the neutron star
surface. We show that such series of bursts are naturally
explained in the model of a spreading layer of accreting matter
over the neutron star surface in the case of a sufficiently high
($\dot{M}\ga 1\times 10^{-9}\ M_{\odot}\ \mbox{yr}^{-1}$)
accretion rate (corresponding to a mean luminosity $L_{\rm
  tot}\ga 1\times 10^{37} \mbox{erg s}^{-1}$). The existence of
triple bursts requires some refinement of the model --- the
importance of a central ring zone is shown. In the standard
model of a spreading layer no infall of matter in this zone is
believed to occur.

\noindent
{\bf DOI:} 10.1134/S106377371709002X


\vfill
\noindent\rule{8cm}{1pt}\\
{$^*$ e-mail $<$sergei@hea.iki.rssi.ru$>$}

\clearpage

\section*{INTRODUCTION}
\noindent
In the period of the discovery of type I X-ray bursts (Belian et
al. 1972, 1976; Babushkina et al. 1975; Grindlay et al. 1976;
Heise et al. 1976) and their theoretical explanation as
thermonuclear explosions of a mixture of hydrogen and helium on
the surface of a neutron star with a weak magnetic field (Hansen
and van Horn 1975; Woosley and Taam 1976; Maraschi and Cavaliere
1977), the question of precisely where the explosion occurred
was not discussed specially. It was believed that the accreting
matter spreads rapidly over the entire neutron star surface, and
the explosion could begin in a more or less arbitrary place in
which critical conditions favorable for this were accidentally
created at a given time. But, most importantly, it was believed
that after the initiation of an explosion the thermonuclear
burning propagates over the entire stellar surface in fractions
of a second with a supersonic (detonation wave) speed, $v_{\rm
  det}\sim10^4\ \mbox{km s}^{-1}\gg
a_{s}=({2kT}/{m_p})^{1/2}\simeq1200\ ({kT}/{10\ \mbox{keV}})^{1/2}\ \mbox{km
  s}^{-1}$; therefore, the specific place of ignition is of no
importance (Joss et al. 1978; Fryxell and Woosley 1982a). Here,
$T$ is the temperature at the base of the neutron star
atmosphere, $a_s$ is the corresponding sound speed, and $m_p$ is
the proton mass.  Except for the short ($\leq1$ s) X-ray flux
growth stage, the observed exponentially decaying burst time
profile is determined by the cooling time of the
explosion-heated stellar atmosphere --- the diffusion time
($\simeq5-15$~s) of X-ray photons undergoing multiple Compton
scatterings in it and free-free production-absorption (Paczinski
1983a, 1983b; Ebisuzaki et al. 1984; Ebisuzaki 1987; Sunyaev and
Titarchuk 1986; London et al. 1986).

Subsequently, however, it became clear that the detonation wave
in the neutron star atmosphere should rapidly decay and could
not ignite an appreciable area of its surface (Timmes and
Niemeyer 2000).  The deflagration wave propagates too slowly
(Fryxell and Woosley 1982b; Nozakura et al. 1984; Bildsten
1995), with a speed $v_{\rm def}\sim 0.01-0.1\ \mbox{km s}^{-1}$
(for conductive energy transfer from the burning region to the
surrounding matter) or $\sim 0.3-3\ \mbox{km s}^{-1}$ (for
convective energy transfer). In this case, the flame propagation
time is much longer than the observed burst duration. There were
doubts that the explosion was capable of affecting the entire
surface of the star.

The conclusion about fairly slow flame propagation was confirmed
by the RXTE discovery of high-frequency (ms) coherent
oscillations of the X-ray flux from bursters during bursts,
which were explained by the neutron star spin with a frequency
$\Omega_s\sim300-600$ Hz (Strohmayer et al. 1996, 1997; Smith et
al. 1997; Galloway et al. 2008). The flame speed $v$ must be
less than $\pi R_* \Omega_s\,/N\sim 150 \Omega_s/(\mbox{400
  Hz})\ \mbox{km s}^{-1}\ll a_s$, where $R_*$ is the neutron
star radius, which is assumed here and below to be 12 km, and
$N\ga100$ is the number of successive pulsations needed for
their reliable detection ($T=N/\Omega_s\ga 0.25$ s is the
detection period). However, it is apparently still higher than
the deflagration speed $v_{\rm def}$.

Quite recently, three-dimensional theoretical computations of an
explosion on the surface of a neutron star (Simonenko et
al. 2012; see also Gryaznykh 2013a, 2013b) have shown that the
flame can propagate in a qualitatively different,
three-dimensional way.  These authors pointed out that the heat
flux from the explosion along the stellar surface should weaken
greatly, because the bulk of the explosion energy (radiative and
kinetic) is carried away upward due to the atmosphere being
exponential. The horizontal heat flux may turn out to be
insufficient for direct ignition of the surrounding matter. In
these conditions the flame will propagate through the inflow of
matter expanding during the explosion onto the layer of the
unperturbed atmosphere surrounding the place of explosion, its
pressing and, thus, the creation of conditions for thermonuclear
ignition at the base of the atmosphere. The speed of this
process $v_{\rm 3D}$ can exceed appreciably the deflagration
speed $v_{\rm def}$, while during powerful explosions it can
reach and even exceed the sound speed $a_s$. Such a process
allows the observed duration of ordinary bursts to be explained.

In this paper we discuss the detection of series of multiple
(triple and double) type I X-ray bursts from a number of known
bursters occurring with a recurrence time $t_{r}\sim 400$--$600$
s ($7$--$10$ min) in the data of JEM-X telescope (Lund et al.
2003) onboard the INTEGRAL observatory (Winkler et
al. 2003). This time is much shorter than the characteristic
time of matter accumulation $t_{a}\sim 4\pi \Sigma_c R_*^2
\dot{M}^{-1}\simeq 5\ \Sigma_8\, \dot{M}_{17}^{-1}$ h needed for
a thermonuclear explosion to be initiated on the neutron star
surface. Here, $\Sigma_c=10^8\, \Sigma_{8}\ \mbox{g cm}^{-2}$ is
the critical surface density of the accumulated matter (for
explosive helium ignition), $\dot{M}=10^{17}\,
\dot{M}_{17}\ \mbox{g s}^{-1}$ is the accretion rate
corresponding to the total luminosity of the neutron star in the
period between bursts $L_{X}=GM_*\dot{M}/R_*\simeq 1.6\times
10^{37}\ \dot{M}_{17}\ \mbox{erg s}^{-1}$, $G$ is the
gravitational constant, and $M_*$ is the neutron star mass,
which is assumed to be $1.4\ M_{\odot}$.  The time $t_{a}$
determines the mean frequency of ordinary bursts $<\nu>=1/t_{a}$
observed from each specific burster, and, of course, most of the
bursts from the bursters being discussed were detected precisely
with this frequency. The detection of multiple bursts recurring
on a time scale $t_{r}\sim 10\ \mbox{min}\ll t_{a}$ from the
same bursters seems surprising.

Multiple bursts were also detected previously, in particular,
double bursts were observed by Murakami et al. (1980) and Ohashi
et al. (1982) from the sources 4U\,1608-522 and 4U\,1636-536,
more rare triple ones were observed by Boirin et al. (2007) from
the transient burster EXO\,0748-676 and by Keek et al. (2010)
from several more sources. Sanchez-Fernandez et al. (2011)
detected a triple burst from the burster 4U\,1608-522 with the
JEM-X telescope onboard the INTEGRAL observatory, the same one
whose data are analyzed in this paper. In addition to bursts
with a recurrence time $t_{r} \sim 10$ min, double bursts with
$t_{r} \sim 10$ s were detected. Bhattacharyya, Strohmayer
(2006) observed such a burst from the source 4U\,1636-536; they
assure that the burst profile was formed without involving any
photospheric expansion effects.

The nature of multiple bursts is not yet clear.  Having studied
the properties of such bursts detected during superlong (158 h)
continuous observations of the burster EXO\,0748-676 by the
XMM-Newton satellite, Boirin et al. (2007) revealed: (1) a
reduction in the exponential decay time of the profile for the
recurrent events compared to the initial ones (which, in their
opinion, is related to a decrease of the hydrogen abundance in
the exploded matter) and (2) a decrease in the intensity of
events in the series compared to the separate bursts from this
source. Based on these properties of multiple bursts, they
assumed that stepwise burning of the layers of stratified
material with different chemical compositions (hydrogen, helium,
and CNO abundances) is observed on the neutron star surface
during them. Hydrogen burns out during the first flash,
whereupon conditions for the ignition of helium are created
etc. Indeed, the computations by Fujimoto et al. (1981) and Peng
et al. (2007) show that at a sufficiently low accretion rate the
hydrogen burning on the neutron star surface can be explosive
and, at the same time, may not accompanied by the simultaneous
ignition of helium. It is only unclear why the helium burning is
resumed $\sim10$ min after the first flare and why triple events
are observed.

In this paper, based on the JEM-X/INTEGRAL observations of
multiple events, we offer a different possible explanation of
this phenomenon: the thermonuclear explosions producing
successive flashes occur in physically separated regions on the
neutron star surface in which the matter efficiently accumulates
during its nonuniform (in the meridional direction) infall as it
is accreted. Such infall of matter should be expected at a
sufficiently high, though much lower than the Eddington level,
accretion rate in the model of a spreading layer proposed by
Inogamov and Sunyaev (1999, 2010).  According to this model,
falling from the accretion disk into the boundary equatorial
region of the star, the accreting matter has an excessively
large (Keplerian) angular momentum, that completely compensates
the gravitational attraction, which together with the radiation
pressure does not allow it to immediately settle to the stellar
surface. Continuing to rotate, the matter is displaced toward
higher latitudes and only there, often in the immediate vicinity
of the neutron star poles, does it slow down and settle to the
stellar surface. Depending on the accretion rate, the ring
regions where the infall of matter occurs can be at different
distances from the equator and can have different widths. The
bulk of the accretion energy of the infalling matter accounted
for by the spreading layer is released and radiated in the zones
of these regions more distant
from the equator. Note that in the Newtonian approximation the
luminosity of the spreading layer is equal to the luminosity of
the entire accretion disk $L_b\simeq0.5
\dot{M}R_*^2(\Omega_K-\Omega_s)^2\simeq 0.5GM_*\dot{M}/R_*$
(Shakura and Sunyaev 1988; Kluzhniak 1988). Here,
$\Omega_K=(GM_*/R_*^3)^{1/2}$ is the Keplerian angular velocity
near the stellar surface.

Suppose that the subsequent spreading of the fallen matter from
these regions over the stellar surface is much slower than its
accumulation, so that a critical surface density $\Sigma_c$ is
reached at some time. The amount of matter fallen in the
northern and southern regions must be approximately the same.
However, it is obvious that the explosion initially begins in
one of them, let this be the northern region.  After the
explosive burnout of hydrogen and helium in it (for example, in
the regime proposed by Simonenko et al. 2012), the flame slowly
(as a deflagration wave) propagates over a less dense layer of
matter with a speed $v_{\rm def}\sim 0.01\ \mbox{km s}^{-1}$
until it reaches the boundary of the southern region, where a
new flash begins. Let us examine how well this explanation
agrees with the observed picture of the series of multiple
bursts.

\begin{figure}[tp]


\epsfxsize=1.0\textwidth
\epsffile{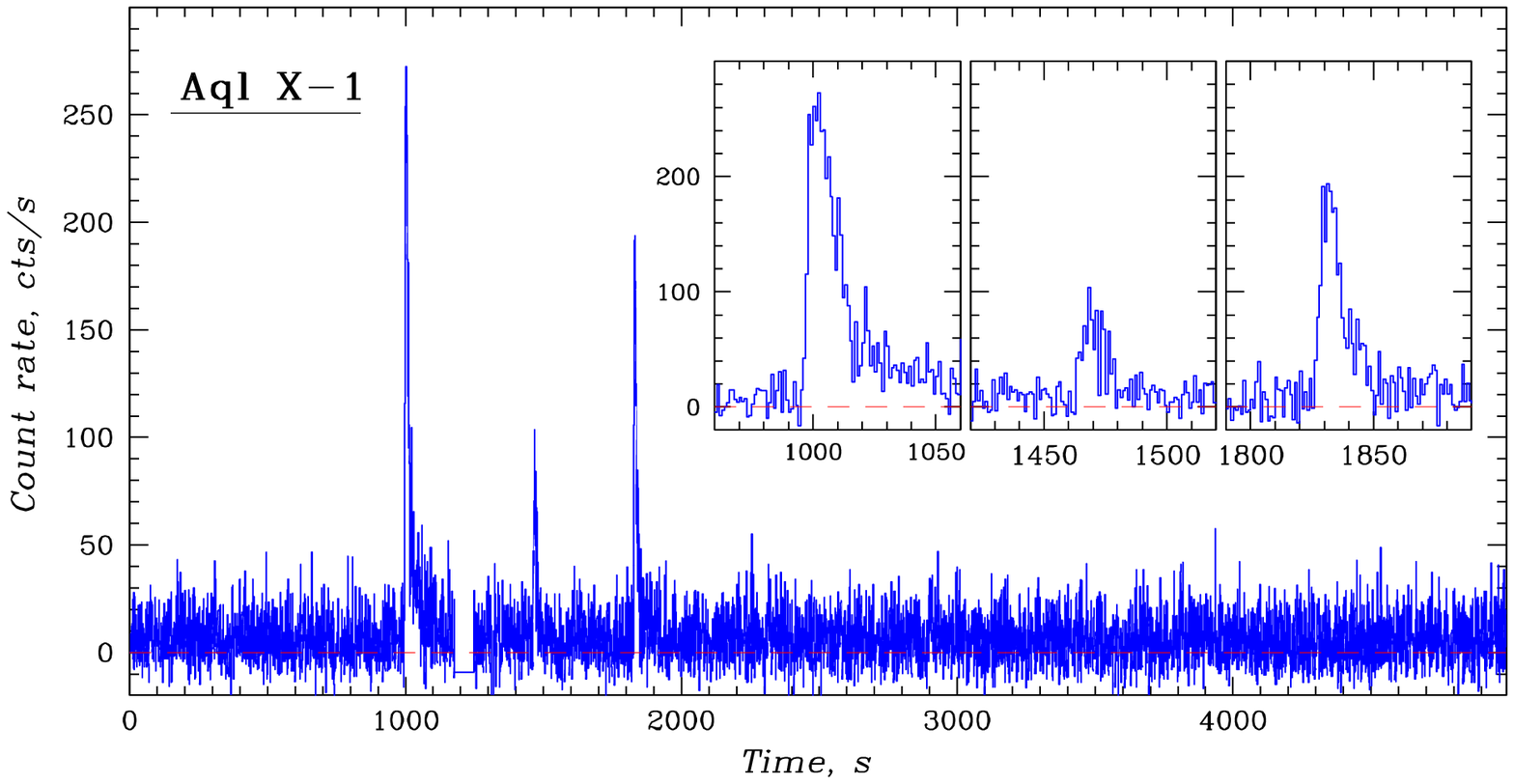}
\epsfxsize=1.0\textwidth
\epsffile{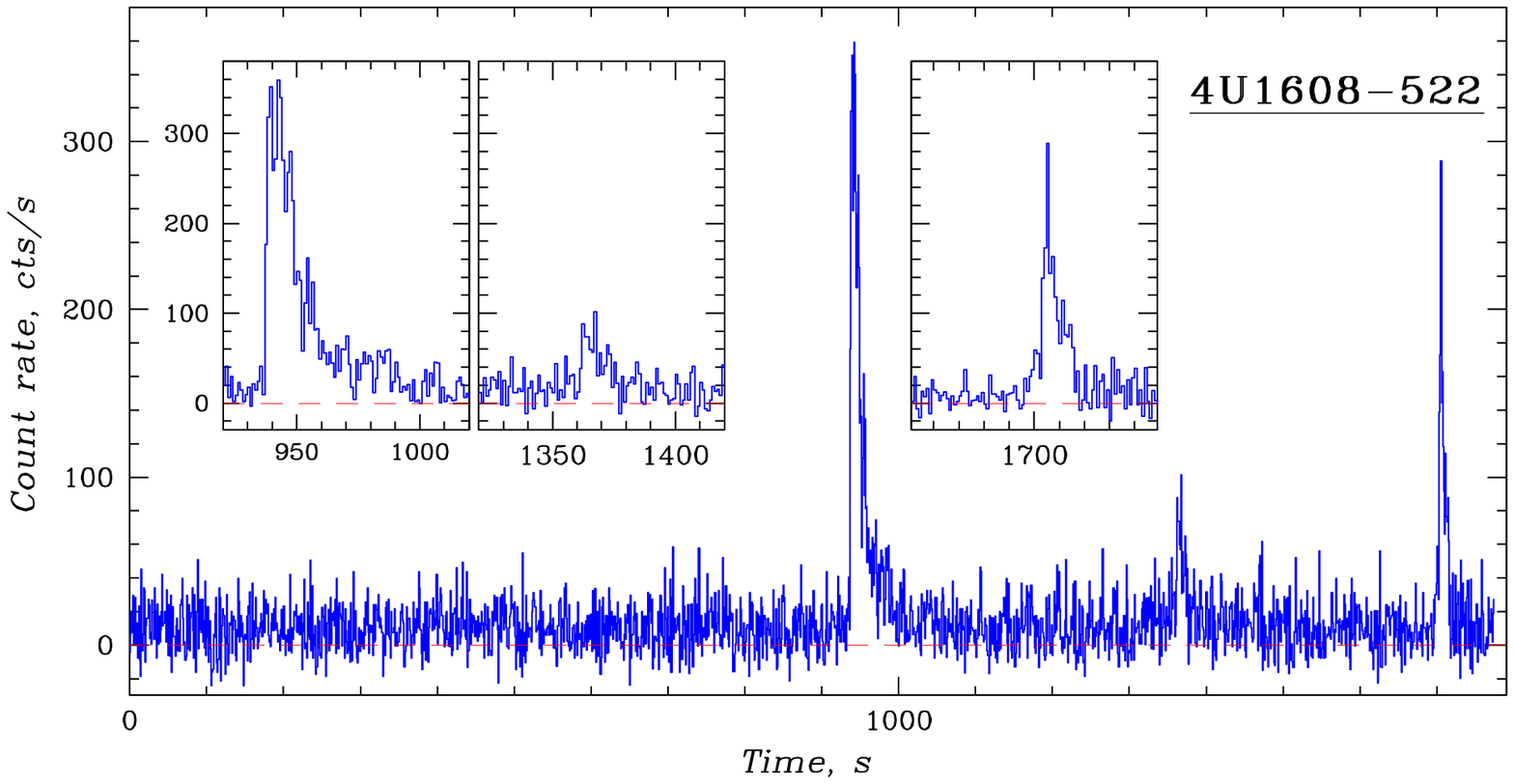}
\caption{\rm JEM-X/INTEGRAL photon count rates (the 3--20 keV
  energy band) in the observing sessions on March 24, 2004
  (top), and March 28, 2009 (bottom), during which triple
  thermonuclear X-ray bursts were detected from the X-ray
  bursters Aql\,X-1 and 4U\,1608-522.  In the inserts the burst
  profiles are shown with a better time resolution.}
\end{figure}

\hspace{-3cm}\begin{table}[p]

\centering {Characteristics  of multiple type I X-ray bursts
  detected by the \mbox{JEM-X} telescope onboard the INTEGRAL
  observatory in 2003--2014} 

\vspace{5mm}
\footnotesize
\begin{tabular}{l|c|c|c|c|c|c|c|c@{\,}|l@{}} \hline\hline
\multicolumn{1}{c|}{Date,\a\ }&Time,\a\ & $\delta T,$\bb\ & $C_{b}(C_{c}),$\cc\ &
$F_{b},$\dd\ & $F_{c},$\e\ &$\Delta T_1,$\f\ &$\Delta T_2,$\g\ &$\Lambda$\h\ &Source\\ [1mm] 
\multicolumn{1}{c|}{UTC}&{UTC}& s & counts& $10^{-8}\ \mbox{erg}$&$10^{-10}\ \mbox{erg}$&h &h & &burst\\ 
&& & s$^{-1}$&$\mbox{s}^{-1}\ \mbox{cm}^{-2}$&$\mbox{s}^{-1}\ \mbox{cm}^{-2}$& & & \\ \hline 
\multicolumn{1}{c|}{1}&2&3&4&5&6&7&8&9&\multicolumn{1}{c}{10}\\ \hline
 
    2004-02-23 & 23\uh20\um34\us& $16$ & $166\ (8)$&$1.6\pm0.9$& $25.2\pm1.0$&  $1.9$& $1.6$&&4U\,1636-536\\
                       & 23\uh47\um57\us& $10$ & $186\ (9)$&$1.5\pm0.9$& & &  &\\ \hline
    2004-03-24 & 17\uh03\um31\us& $19$ & $273\ (7)$&$1.5\pm1.4$& $17.6\pm1.4$&  $830$& $2.6$&&Aql\,X-1\\ 
                       & 17\uh11\um17\us& $  8$ & $104\ (7)$&$1.8\pm1.3$& & &  &\\
                       & 17\uh17\um20\us& $20$ & $194\ (7)$&$3.0\pm1.2$& & &  &\\ \hline
    2004-08-08 & 18\uh30\um11\us& $30$ & $197\ (7)$&$3.0\pm0.8$& $19.4\pm1.5$&  $8.7$& $1.6$&&4U\,1636-536\\
                       & 18\uh50\um34\us& $20$ & $110\ (7)$&$1.0\pm0.6$& & & & \\ \hline
    2004-08-22 & 23\uh52\um34\us& $7$ & $299\ (33)$&$3.5\pm2.2$& $67.8\pm3.2$&  $7.6$& $2.6$&*&GX\,3+1\\
    2004-08-23 & 01\uh03\um12\us&   $4$ & $158\ (30)$&$3.7\pm1.8$&                       & && \\ \hline
    2005-04-15 & 03\uh37\um46\us& $6$ & $289\ (28)$&$4.2\pm1.8$& $66.8\pm2.1$&  $15.3$& $2.6$&*&GX\,3+1\\
                       & 04\uh37\um00\us&   $6$ & $192\ (23)$&$5.8\pm3.4$& $66.0\pm2.1$& && \\ \hline
    2005-04-26 & 10\uh15\um03\us& $39$ & $256\ (11)$&$3.8\pm1.9$& $23.2\pm1.3$&  $11.0$& $2.6$&&Aql\,X-1\\
                       & 10\uh28\um16\us& $14$ & $149\ (11)$&$1.5\pm0.9$& & && \\ \hline
    2005-04-29 & 09\uh00\um36\us& $22$ & $263\ (7)$&$5.2\pm1.5$& $21.6\pm1.3$&  $36.1$& $2.6$&&Aql\,X-1\\
                       & 09\uh11\um23\us& $17$ & $256\ (7)$&$4.7\pm1.3$& & && \\ \hline
    2005-08-26 & 13\uh32\um51\us& $10$ & $201\ (8)$&$16.9\pm8.2$& $26.7\pm1.4$& $1.9$& $1.6$&&4U\,1636-536\\
                       & 13\uh55\um46\us& $9$ & $121\ (8)$&$2.6\pm1.8$& & &  &\\ \hline
    2005-08-27 & 17\uh40\um25\us& $13$ & $144\ (8)$&$1.8\pm1.3$& $18.6\pm1.0$&  $26.2$& $1.6$&&4U\,1636-536\\
                       & 17\uh46\um48\us& $15$ & $160\ (8)$&$1.6\pm1.2$& & && \\ \hline
    2005-10-10 & 23\uh54\um04\us& $9$ & $155\ (5)$&$3.4\pm1.6$& $4.4\pm1.9$&  $94$& $16.5$&*& XTE\,J1739-285\\
     2005-10-11 & 00\uh36\um45\us& $10$ & $100\ (5)$&$1.6\pm0.9$& & &  &\\ \hline
    2006-03-12 & 23\uh26\um54\us& $14$ & $196\ (6)$&$1.5\pm1.1$& $5.6\pm1.8$&  $19.6$& $11.8$&&SAX\,J17470-2853\\
                       & 23\uh40\um00\us& $14$ & $145\ (6)$&$2.9\pm0.8$& & & & \\ \hline 
    2006-03-21 & 21\uh34\um48\us&    $6$ & $157\ (7)$&$2.6\pm1.6$& $3.6\pm2.4$&  $41$& $16.5$&& XTE\,J1739-2885\\
                      & 21\uh40\um03\us&    $4$ & $161\ (7)$&$2.7\pm1.9$& & &  &\\ \hline
    2009-03-28 & 20\uh10\um38\us& $24$ & $359\ (13)$&$7.9\pm3.1$& $30.7\pm1.7$&  $221$& $1.6$&&4U\,1608-522\\
                       & 20\uh17\um43\us& $12$  & $101\ (13)$&$1.5\pm1.4$& & &  &\\
                       & 20\uh23\um21\us& $12$  & $289\ (13)$&$2.8\pm1.5$& & & &  \\ \hline                                                                        
    2010-02-24 & 23\uh35\um49\us& $21$ & $135\ (6)$&$1.7\pm1.1$& $17.9\pm1.8$&  $5.0$& $1.6$&&4U\,1636-536\\
                       & 23\uh53\um41\us& $13$  & $116\ (6)$&$1.4\pm1.0$& $18.6\pm1.7$&  &  &\\ \hline
    2010-04-02 & 11\uh22\um57\us& $11$ & $305\ (31)$&$6.7\pm3.0$& $72.3\pm3.1$&  $86$& $2.6$&*& GX\,3+1\\
                       & 12\uh03\um57\us&   $4$ &   $170\ (33)$&$3.0\pm1.0$& $78.2\pm4.2$& &  &\\ \hline
    2010-04-13 & 15\uh47\um07\us&  $7$  & $283\ (29)$&$4.7\pm4.0$& $68.9\pm4.6$&  $130$& $2.6$&*& GX\,3+1\\
                       & 16\uh28\um50\us&  $6$  &$214\ (34)$&$3.7\pm1.4$& $68.9\pm4.6$& &  &\\ 
                       & 17\uh03\um25\us&  $4$  &$130\ (29)$&$2.6\pm0.9$& $69.9\pm4.3$& &  &\\ \hline         
\multicolumn{10}{l}{}\\ [-3mm]
\multicolumn{10}{l}{\a\ Date and time (UTC) of the peak count
  rate in the burst.}\\
\multicolumn{10}{l}{\bb\ The burst duration.}\\
\multicolumn{10}{l}{\cc\ The peak $C_b$ and mean $C_c$ count rate
  of photons  from the source in the 3--20 keV band.}\\
\multicolumn{10}{l}{\dd\ The peak flux in the 3--20 keV band.}\\
\multicolumn{10}{l}{\e\ The persistant flux from the burster
  during several previous exposures in the 3--100 keV band}\\
\multicolumn{10}{l}{\f\ The time from the first burst in the
  series to the nearest burst not from the series $\Delta T_1$.}\\
\multicolumn{10}{l}{\g\ The minimum time between ordinary bursts
  from this burster for the entire sample $\Delta T_2$.}\\
\multicolumn{10}{l}{\h\ The interval between the bursts in the series excceeds 30 min.}\\
\end{tabular}
\end{table}

\section*{OBSERVATIONS}
\noindent
We revealed multiple events when working on the full catalog of
X-ray bursts detected by the \mbox{JEM-X} telescope onboard the
INTEGRAL observatory in 2003-2014 (Chelovekov et al. 2017). This
is the final, third part of our investigation of thermonuclear
X-ray bursts with the INTEGRAL telescopes.  The first two parts
(Chelovekov at al. 2006; Chelovekov and Grebenev 2011) were the
catalogs of bursts detected by the IBIS/ISGRI telescope
sensitive in a harder X-ray band ($\ga18$ keV) than the JEM-X
band (4--30 keV).  Although it is clear that it is better to
observe type I bursts in the standard X-ray band, the search for
bursts in the IBIS/ISGRI data made sense due to the larger field
of view of this telescope (exceeding the JEM-X field of view by
a factor of $\sim 8$) and better corresponded to the main goal
of our investigation, i.e., revealing hitherto unknown
bursters. Two such bursters have indeed been discovered (see
Chelovekov and Grebenev 2007, 2010).

The full catalogue of bursts detected with JEM-X is accessible
at {\sl $<$http:\,//dlc.rsdc.rssi.ru$>$}. The main characteristics
of the series of double and triple events selected from this
catalog are given in the table. It provides the date of
observation, the time of the peak count rate $T$ and the
duration $\delta T$ of each burst in the series, the peak and
observation-averaged ($\sim1$ h) count rates of events from the
source $C_b$ and $C_c$, the recorded peak flux in the burst
$F_b$, and the mean flux $F_c$ from the source in several
consecutive previous exposures (for more details, see Chelovekov
et al. 2017). The fluxes in the bursts were measured in the
3--20 keV band; the persistent flux was measured in the 3--100
keV band. Note that the characteristic persistent flux from the
bursters being discussed $F_c\sim 1\times 10^{-9}\ \mbox{erg
  s}^{-1} \mbox{cm}^{-2}$ corresponds to an X-ray luminosity
$L_X\sim 1.2\times 10^{37}\ \mbox{erg s}^{-1}$ under the
assumption that the source is near the Galactic center at a
distance of 8 kpc, i.e. these are all sources with a fairly high
accretion rate.
\begin{figure}[tp]


\epsfxsize=1.0\textwidth
\epsffile{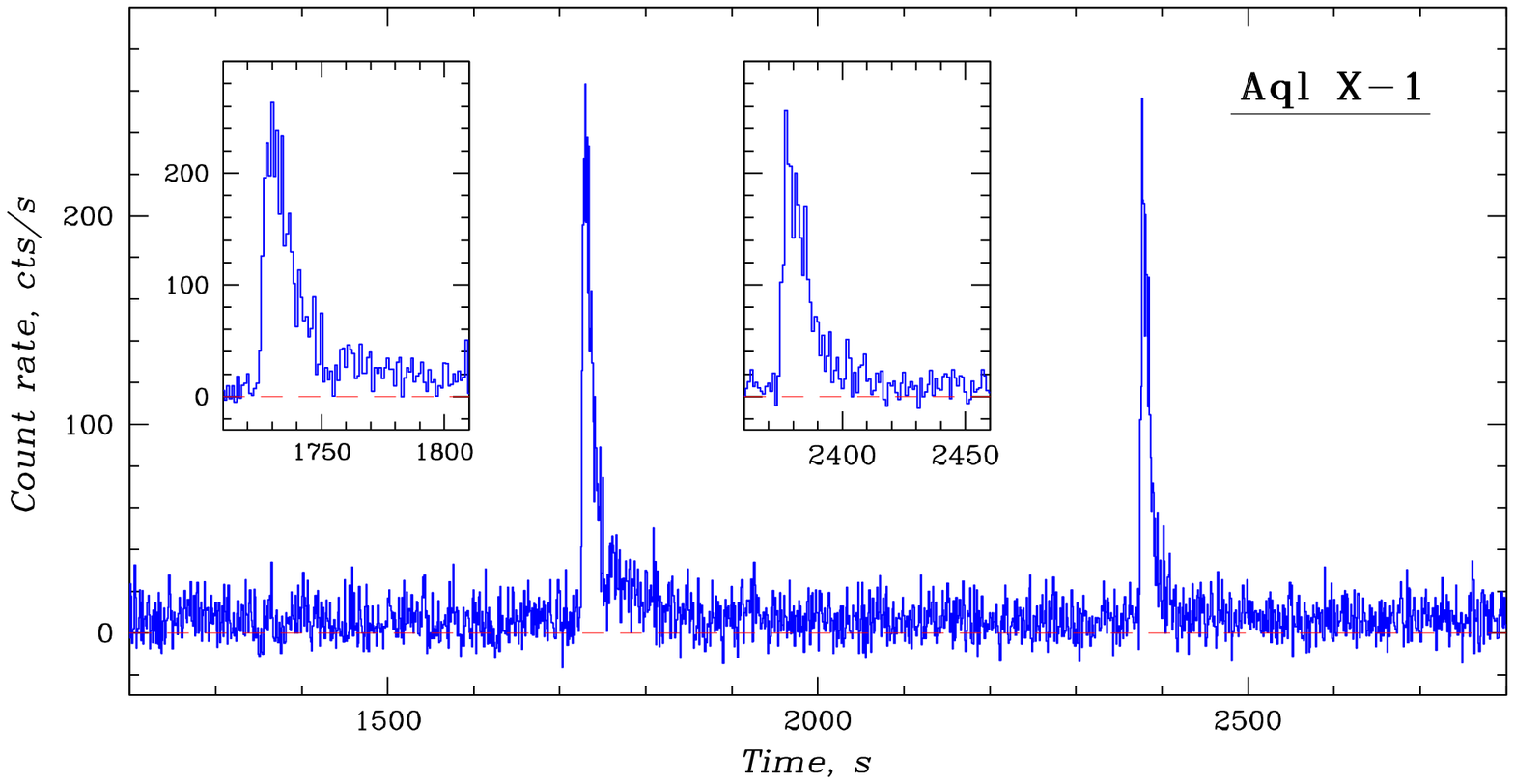}
\epsfxsize=1.0\textwidth
\epsffile{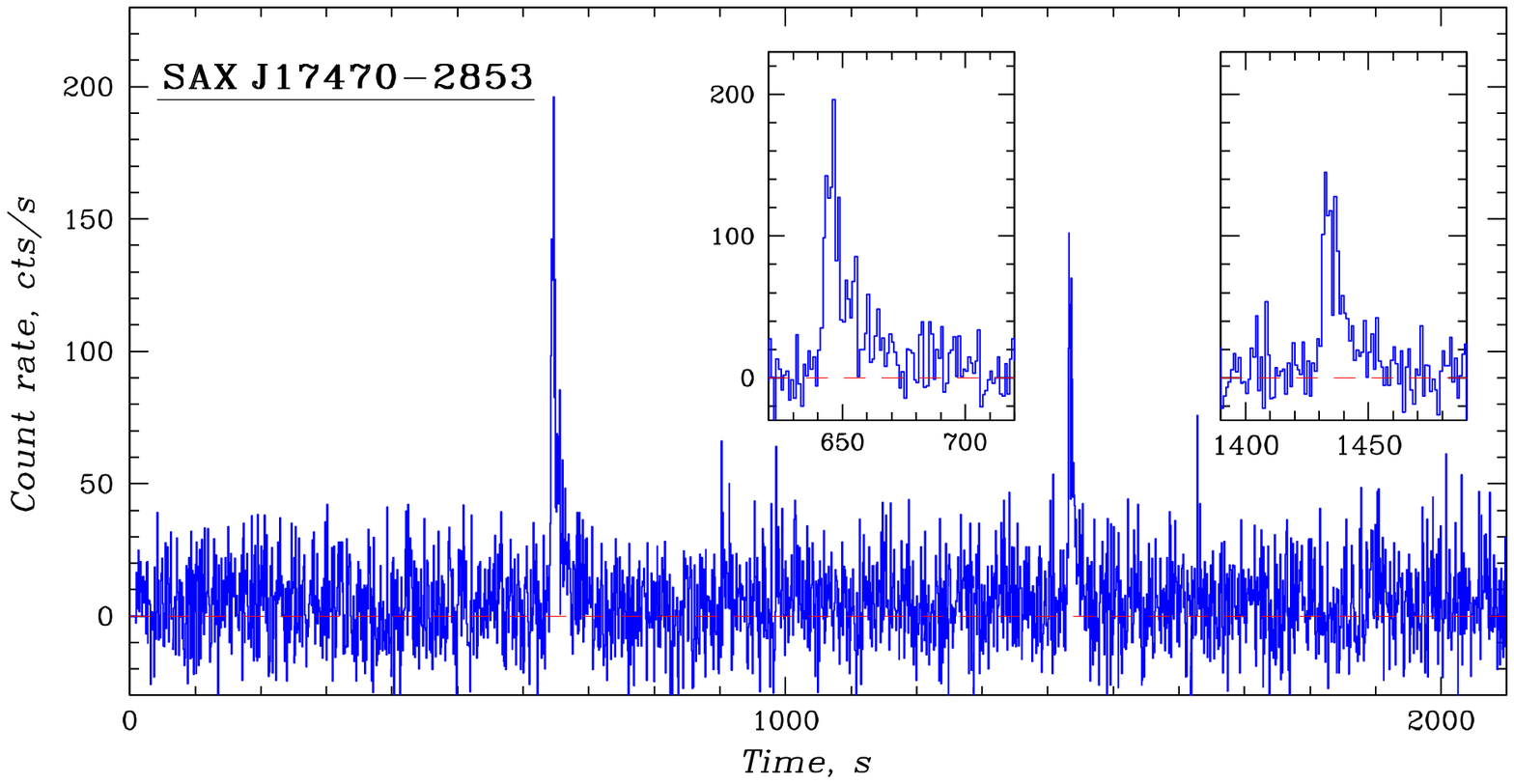}

\caption{\rm JEM-X/INTEGRAL photon count rates (the 3--20 keV
  energy band) in the observing sessions on April 29, 2005
  (top), and March 12, 2006 (bottom), during which double
  thermonuclear X-ray bursts were detected from the X-ray
  bursters Aql\,X-1 and SAX\,J17470-2853.  In the inserts the
  burst profiles are shown with a better time resolution.}
\end{figure}

The data from the table allow us to estimate the mean time
interval between single bursts (or a single burst and a series
of bursts) $t_a\simeq 0.24 \alpha \Sigma (\delta\, T F_b)\,
/\,F_c$ using the so-called parameter $\alpha\sim 40$ (Lewin et
al. 1993), which characterizes the efficiency of energy release
during accretion compared to explosive helium burning. Here, the
factor 0.24 allows for the deviation of the measured burst
duration $\delta T$ from the exponential time and $\Sigma$
denotes summation over the bursts of the series. In particular,
for the known bursters Aql X-1 and 4U\,1608-522 $t_a\simeq
1.6-2.4$ h .  The table provides the time interval $\Delta T_1$
from the first burst in the series to the nearest burst from
this burster not from this series and the minimum interval
$\Delta T_2$ between ordinary bursts from this burster for the
entire sample. These intervals allow one to judge the mean
frequency of bursts $<\nu>=t_a^{-1}$ from a given burster at the
current and mean accretion rates, respectively. Since the
INTEGRAL observations of each specific source were generally
episodic, though they could last tens of hours, some bursts must
have undoubtedly been missed; therefore, these intervals should
be considered only as upper limits on the time $t_a$. For this
reason, we, in particular, used a fairly stringent criterion for
the inclusion of bursts in the series: the interval between them
should not exceed 30 min. The asterisks in column 9 of the table
mark several possible double bursts that did not pass this
criterion. These were detected from the known bursters GX\,3+1
and XTE\,J1739-285; the interval between them was 40--70 min,
while the time $\Delta T_2$ for these sources was 2.6 and 16.5
h, respectively. It is unclear whether these bursts are events
similar to the remaining multiple events in the table and their
long recurrence time $t_r$ reflects some of their unique
physical properties or these are ordinary single bursts
distinguished by an unusually short accumulation time $t_a$ of
the critical matter density. Estimates based on the data from
the table similar to those given above for the bursters Aql~X-1
and 4U\,1608-522 show that this is possible at least for the
source GX\,3+1. On the other hand, in this case, it is most
likely insufficient for the source to have an enhanced accretion
rate compared to other bursters; it is also necessary that at
such an accretion rate the explosive development of a flash does
not pass into continuous burning (see, e.g., Strohmayer and
Bildsten 2006).
\begin{figure}[tp]


\epsfxsize=1.0\textwidth
\epsffile{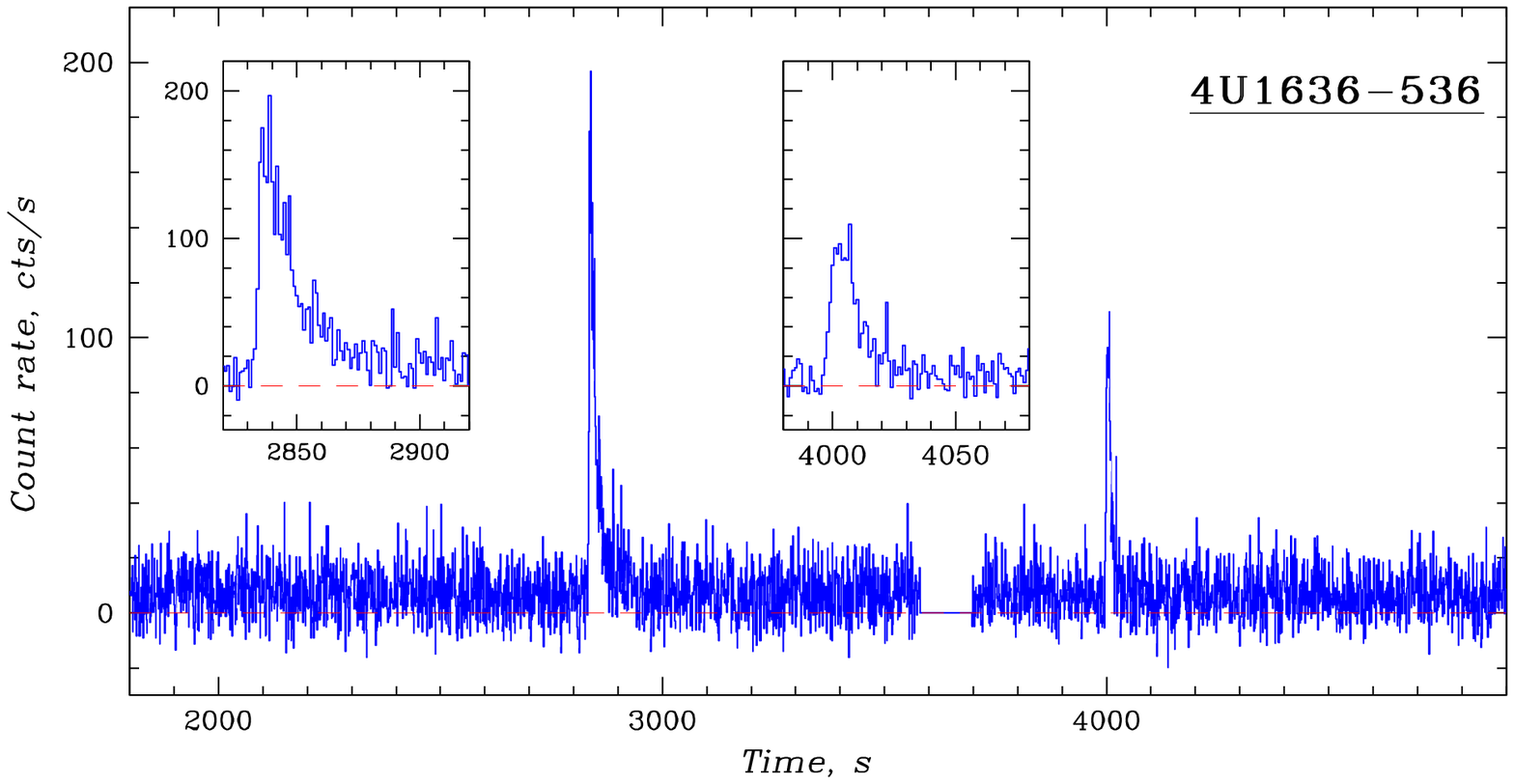}
\epsfxsize=1.0\textwidth
\epsffile{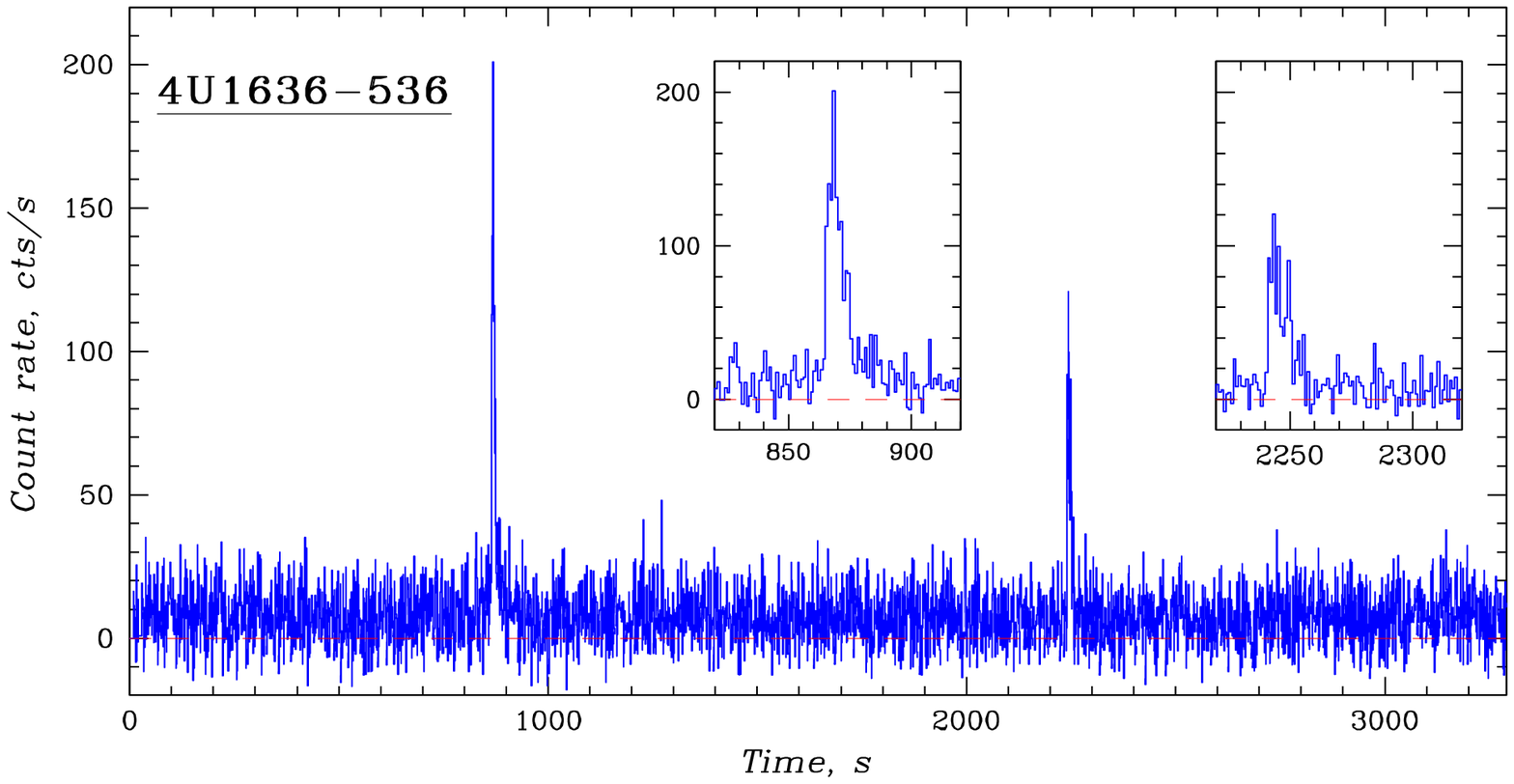}

\caption{\rm JEM-X/INTEGRAL photon count rates (the 3--20 keV
  energy band) in the observing sessions on August 8, 2004
  (top), and August 26, 2005 (bottom), during which double
  thermonuclear X-ray bursts were detected from the X-ray
  burster 4U\,1636-536. In the inserts the burst profiles are
  shown with a better time resolution.}.
\end{figure}
\begin{figure}[tp]


\epsfxsize=1.0\textwidth
\epsffile{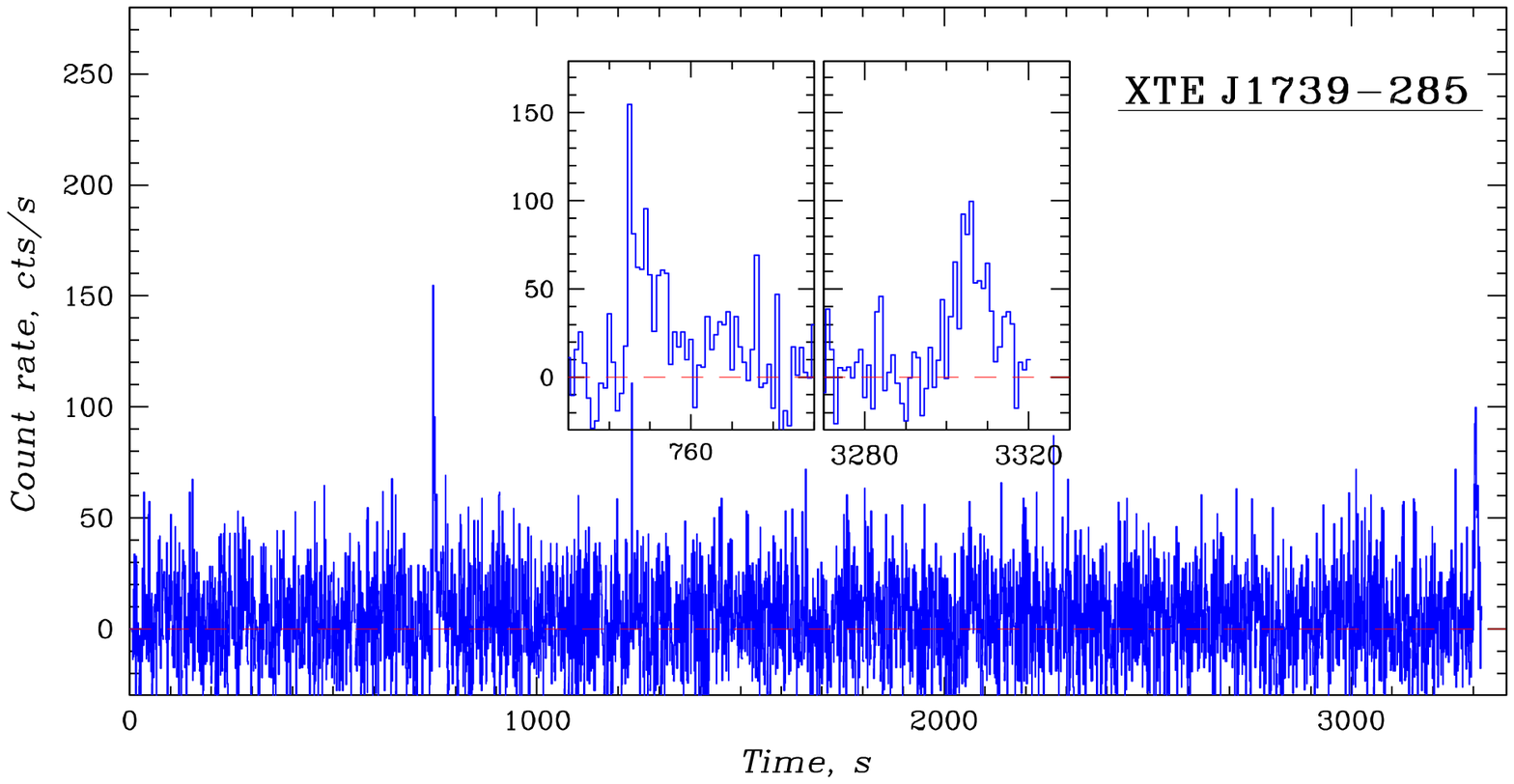}
\epsfxsize=1.0\textwidth
\epsffile{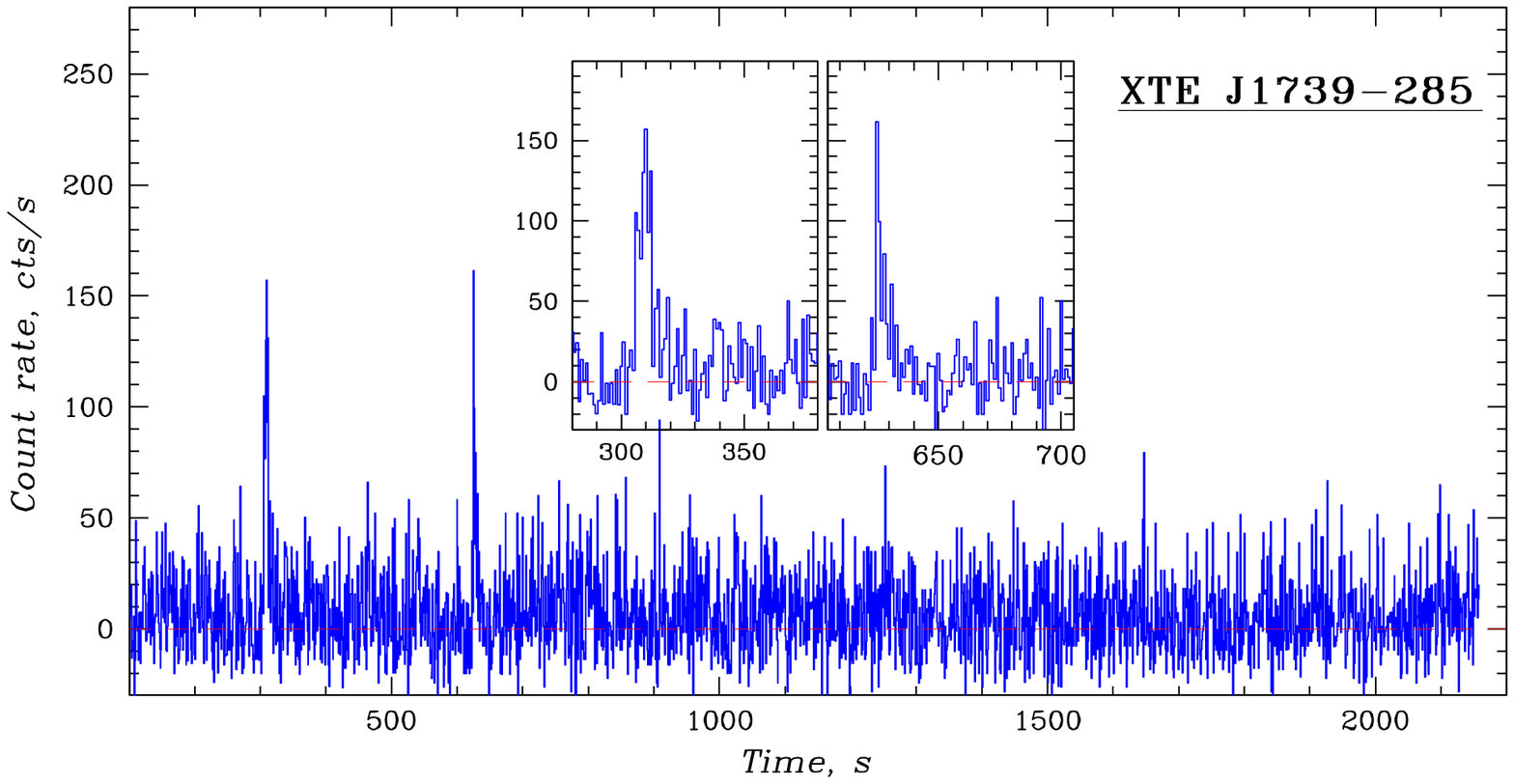}

\caption{\rm  JEM-X/INTEGRAL photon count rates (the 3--20 keV
  energy band) in the observing sessions on October 10--11, 2005
  (top), and March 21, 2006 (bottom), during which double
  thermonuclear X-ray bursts were detected from the X-ray
  burster XTE\,J1739-285. In the inserts the burst profiles are
  shown with a better time resolution.}.
\end{figure}
\begin{figure}[tp]


\epsfxsize=1.0\textwidth
\epsffile{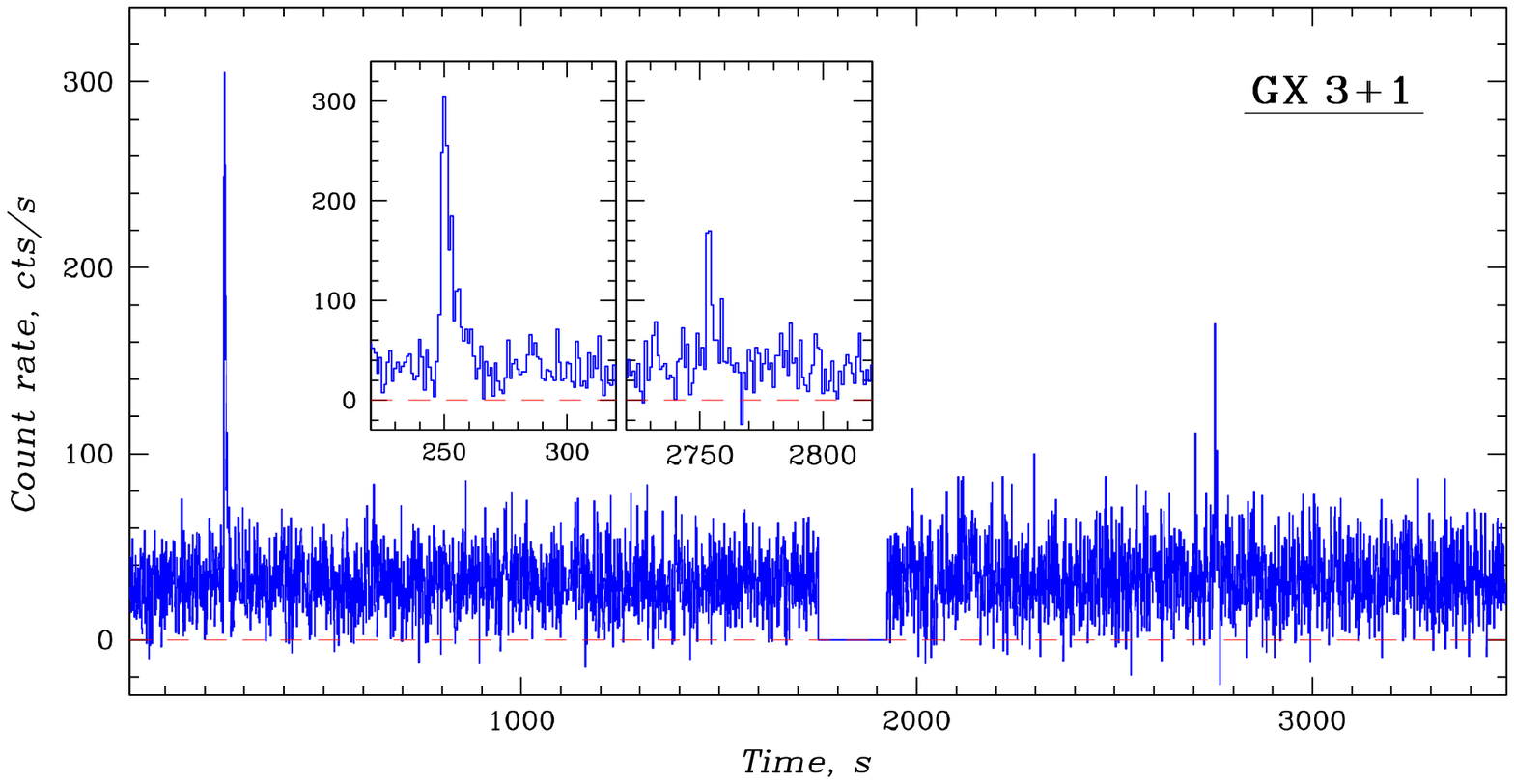}
\epsfxsize=1.0\textwidth
\epsffile{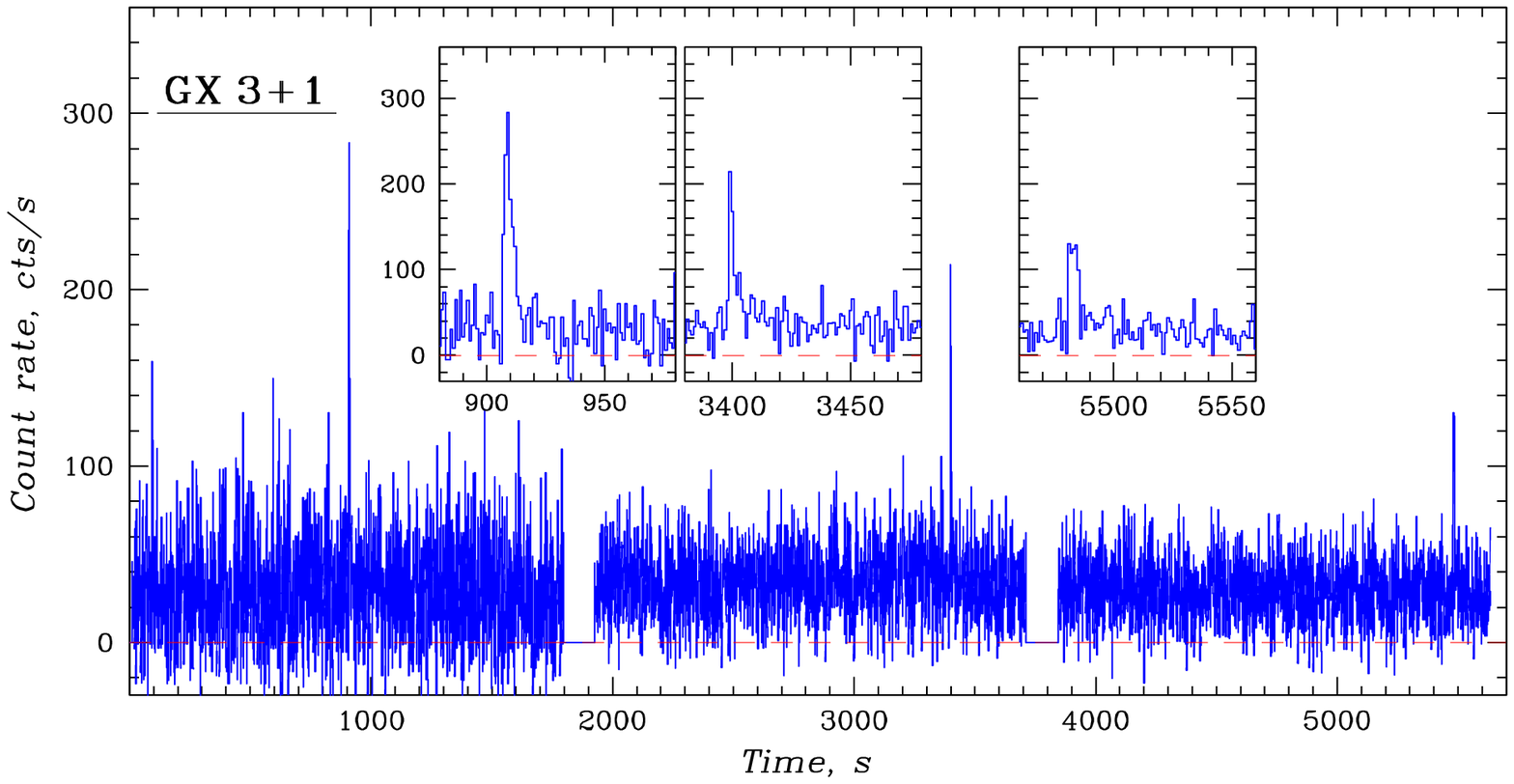}

\caption{\rm JEM-X/INTEGRAL photon count rates (the 3--20 keV
  energy band) from the source GX\,3+1 in the observing sessions
  on April 2 (top) and 13 (bottom), 2010, during which double
  and triple thermonuclear X-ray bursts were detected by this
  telescope. The time interval between the bursts is shorter
  than the sample-averaged interval between the bursts from this
  source only by a few times. In the inserts the burst profiles
  are shown with a better time resolution.}.
\end{figure}

As an example, Fig.\,1 shows the light curves of the triple
events observed from the known bursters Aql~X-1 and
4U\,1608-522; Figs.\,2--4 show those of the double events
observed from the mentioned burster Aql~X-1 and the equally
known bursters SAX\,J17470-2853, 4U\,1636-536, and
XTE\,J1739-285. The insets present the profiles of individual
bursts with a better time resolution demonstrating a fast rise
to the maximum and a long exponential decay, which is
characteristic for type I (thermonuclear) bursts. The burst
duration varied between $\sim5$ s and $\sim 15$ s, i.e., it was
also typical for such bursts.  In all the observed cases of
triple events, the second burst was noticeably fainter than the
first and third bursts and was located almost halfway between
them (closer to the third burst by $\sim50-70$ s). The third
burst was slightly fainter than the first one. The total
duration of the shown series of triple events was $\sim800$ s
($\sim13$ min). In the double events in Figs.\,2--4 the second
burst was, as a rule, fainter than the first one; in some cases,
the bursts had a comparable intensity. As a rule, the duration
of the second burst was shorter than that of the first one.  For
a number of sources the time interval between the double bursts
was comparable to the duration of the triple events ($\sim12-13$
min, cf. Figs.\,1 and 2); for others it was noticeably longer
($\sim20-25$ min for the burster 4U\,1636-536, Fig.\,3). It is
important that the duration of the triple and double series of
bursts for the source Aql~X-1 was the same, suggesting that in
the latter case the intermediate burst was too faint to be
detected. The double bursts from 4U\,1636-536 in Fig.\,3, along
with other series of bursts from this source (see the table),
had almost the same duration, always much longer than $13$ min
characteristic for Aql~X-1. This suggests that the series
duration reflects the neutron star properties, the accretion
rate characteristic for a given source, and/or the composition
of the accreting matter. On the other hand, both anomalously
long ($\sim 42$ min, Fig.\,4a) and anomalously short ($\sim 5$
min, Fig.\,4b) series of two bursts were detected for the source
XTE\,1739-285.

For one more source, the known burster GX~3+1, all the detected
series of bursts were anomalously long, $\ga40$ min
(Fig.\,5). The recurrence time of ordinary bursts from this
source averaged over the entire sample of events recorded by the
\mbox{JEM-X} telescope, $t_a$, was longer only by a factor of
$\sim4$ (see the table). Therefore, as has already been noted
above, we cannot unambiguously assert that series of bursts
similar to those discussed above rather than pairs of close
single bursts were actually detected. Besides, this source is
characterized by a very high persistent flux $(65-80)\times
10^{-10}\ \mbox{erg s}^{-1} \mbox{cm}^{-2}$. It exceeds the
fluxes from other bursters by several times and clearly suggests
a high accretion rate onto this source. On the other hand, it is
interesting that apart from the double bursts, a triple burst
was also detected from this source (Fig.\,5, bottom panel), whose
properties, except for the time scale on which it developed, are
very close to those of the triple bursts from other bursters
(Fig.\,1). In particular, the middle event in this burst also occurred
with a delay by $\sim15$\% relative to the middle of
the interval between the first and last events.

\section*{DISCUSSION}

\noindent
Our analysis of a sample of multiple X-ray bursts detected by
the \mbox{JEM-X} telescope onboard the INTEGRAL observatory has
revealed several trends that can give a key to understanding
this interesting phenomenon.

\begin{enumerate}
\item The profiles of such bursts and especially triple
  bursts with powerful first and last events and a much fainter
  middle event are unique and uniform for different sources;
  these are very difficult to explain in the model of
  successively resuming thermonuclear burning of stratified
  (consisting of the layers of different elements) fuel.

\item The double bursts can be failed triple bursts in
  which the middle burst was too faint to be detected.
  
\item The duration of the series of bursts is, on
average, unique for each specific source, probably
reflecting the parameters of the neutron star in it, the
characteristic accretion rate, and the composition of
the accreting matter.
  
\item The intermediate (second) burst of the triple
events is delayed relative to the middle of the interval
between the first and last events by 10--15\%.
  
\item In addition to the previously discussed series of
bursts with a total duration $\sim10$ min, there can exist
series of bursts with a duration $\ga40$ min from some
sources.
\end{enumerate}

The unique profiles of multiple bursts are naturally explained
in the model of a spreading layer of accreting matter over the
neutron star surface (Inogamov and Sunyaev 1999, 2010). In this
model, reaching the surface of the neutron star in the
equatorial region, the accretion disk matter is displaced in a
spiral toward its poles and only there does it lose its angular
momentum in two ring zones, radiate the energy being released,
and settle to the surface. The probability of reaching the
critical conditions for thermonuclear ignition of the matter
accumulated during accretion is high precisely in these
regions. Since it is obvious that the flash begins initially
only in one of the ring zones, it will be responsible for the
first most powerful burst in the series. Once the matter
accumulated in this zone has burnt out, the thermonuclear flame
propagates with a deflagration wave speed $v_{\rm
  def}\simeq0.01\ \mbox{km s}^{-1}$ over the stellar surface
toward the equator and then toward the opposite stellar pole and
the second ring zone. On reaching it, the last burst in the
series begins. Note that although the bulk of the matter during
accretion falls in these ring zones, some moderate amount of
matter must also settle from the spreading layer on its way to
these zones; otherwise there would be no radiative energy for
its maintenance (recall that the layer must be a
radiation-dominated and levitating one). It is through this
settled matter that the deflagration wave propagates after the
first explosion. The matter from the ring zones that slowly
spreads over the neutron star surface can also contribute to the
layer of fuel accumulated here.

There are several points that do not seem natural in the model of
a spreading layer and require an explanation.

\subsection*{\it The Origin of the Middle Burst in a Series}
\noindent
One would think that the intermediate (second) burst in a triple
series of bursts cannot be explained in any way in the model of
a spreading layer. Proposing it, Inogamov and Sunyaev (1999)
assumed that all of the matter from the disk spread in
meridional directions.  Actually, this cannot be the case,
because the matter in the disk has quite a distinct radial
velocity $v_r=\dot{M}/(2\pi R_*\Sigma_d),$ that must not be
ignored. This velocity can be slowed down only through viscosity
in a narrow equatorial ring layer like the boundary layer
described by Shakura and Sunyaev (1988, 1999) and Kluzhniak
(1988). In this case, at least for part of the accreting matter,
not only the radial velocity but also the rotation velocity
decreases down to the stellar rotation velocity. This matter
settles to the stellar surface straight in the equatorial
zone. Although the amount of matter settling in this zone and
the energy being released in this case are small compared to the
matter and the energy settling and being released, respectively,
in the polar ring zones of the neutron star, it may turn out to
be sufficient to explain the intermediate burst in series of
triple burst events. If, alternatively, little matter fell in
this zone, then we will be able to see only a double burst. The
infall of matter in this zone will be considered in more detail
in Grebenev (2017).

\subsection*{\it Explaining the Asymmetry of the Profile fot Triple Bursts}
\noindent
It has been noted above that the intermediate burst in the
triple events observed from the bursters Aql~X-1 and
4U\,1608-622 is delayed by $\sim50-70$ s relative to the middle
of the time interval between the first and third bursts. At
first glance this delay contradicts the described symmetric
picture. Note, however, that the ring zones in which the matter
settles and accumulates can be quite extended, depending on the
accretion rate (Inogamov and Sunyaev 1999). The first burst
begins in the region of maximum surface density that can be near
the high-latitude edge of the ring zone. Thereafter, all this
zone (or is sufficiently dense part) will be affected by the
flame in a time comparable to the duration of the first
burst. At the same time, the ignition of the opposite zone
begins from its low-latitude edge by the flame front going away
from the equator. Thus, the third burst will begin earlier than
the first burst relative to the time of passage of the
equatorial zone by the flame front and its ignition (i.e., the
second burst in the series). In principle, the observation of
such triple bursts will allow one to investigate the parameters
of the spreading layer and to check the computations performed
by Inogamov and Sunyaev (1999).

\subsection*{\it Series of Bursts of Greater Multiplicity}
\noindent
Keek et al. (2010) reported the detection of a series of X-ray
bursts from the source 4U\,1636-538 consisting of four
events. Such bursts of greater multiplicity can be explained in
the proposed model by assuming that shortly before their
observation the accretion rate onto the source changed
abruptly. In this case, as the burning front passed over the
neutron star surface, the flashes should have occurred in two
ring zones associated with the infall of matter at the initial
accretion stage and two other ring zones associated with the
infall of matter at the final stage. Clearly, this event
requires the fulfillment of certain conditions and can occur
very rarely, much more rarely than double and triple
bursts. This is generally observed.

\subsection*{\it Why are Single Bursts observed?}
\noindent
Or why are double and triple bursts encountered quite rarely?
The point is that the model of a spreading layer acts only at
sufficiently high accretion rates
$\dot{M}\ga0.01\ \dot{M}_{ed},$ where $\dot{M}_{ed}$ is the
critical Eddington accretion rate (Inogamov and Sunyaev 1999).
As the accretion rate decreases, the ring zones of the main
energy release and the infall of accreting matter narrow down
and are displaced toward low latitudes.  At $\dot{M}\la
0.01\ \dot{M}_{ed}$ the entire matter settles in the equatorial
zone; accordingly, only one X-ray burst is observed during the
thermonuclear explosion in this zone. Moreover, the picture is
complicated by the fact that as the accretion rate increases,
the thermonuclear burning may not be accompanied by an explosion
but be continuous. The exact conditions under which it is
possible to observe multiple bursts can be clarified only in
future, through detailed numerical simulations of the
thermonuclear burning in the physical model being discussed.

\section*{ACKNOWLEDGMENTS}
\noindent
This work is based on the INTEGRAL data retrieved via its
Russian and Europian science data centers. It was financially
supported by the ``Transitional and Explosive Processes in
Astrophysics'' Subprogram of the Basic Research Program P-7 of
the Presidium of the Russian Academy of Sciences, the Program of
the President of the Russian Federation for support of leading
scientific schools (grant NSh-10222.2016.2) and the ``Universe''
theme of the scientific research program of the Space Research
Institute, the Russian Academy of Sciences.

\pagebreak   


\vspace{1cm}

\begin{flushright}
{\it  Translated by V. Astakhov\/}
\end{flushright}
\end{document}